\documentclass[12pt]{article} 
\begin{document}

\title{The ``fingers'' of the physics\\
Les ``doigts'' de la physique}

\author{Luigi Foschini\\
{\small Institute TeSRE--CNR, Via Gobetti 101, I-40129, Bologna (Italy)}\\
{\small Email: \texttt{foschini@tesre.bo.cnr.it}}}

\maketitle

\begin{abstract}
The passage of particles through matter is one of the principal ways to 
investigate nature. In this article, we would like to outline the most 
important stages in the development of the theory about the stopping power.\\

\vskip 10pt
Le passage des particules \`a travers de la mati\`ere c'est une des routes principales
pour \`etudier la nature. In cet article nous voulons tracer les stages principales de
le d\'eveloppement des th\'eorie \`a propos du pouvoir d'arr\^{e}t.
\end{abstract}

\vskip 24pt

\noindent PACS-01.65.+g   History of science.\\
PACS-34.50.Bw Energy loss and stopping power.

\section{Introduction}
To investigate nature man has always used means which could explore the
world beyond human perception. Among these methods, the study of the
passage of particles and radiations through matter has been an
extremely thriving sector and one of basic importance for modern
physics; moreover, it has put to the test the minds of the greatest
scientists of this century: Bohr, Bethe, Fermi, just to quote some of
them, for the list would be too long.  

The knowledge of the
interactions that take place in the passage of particles and radiations
through matter, allowed possible to develop several detectors, on whose
characteristics the success of an experiment may depend or not. From
the analysis of this phenomena, many branches of physics and
astrophysics have originated; for example the study of matter in its
elementary components, the study of cosmic rays, not to speak about the
several applications in the biomedical field.  

The main sector of application of these studies is, without doubt, particles physics. It is
enough to think about the impact that Wilson's chamber had in modern physics: we
cannot observe a particle, but, at least, we can see the effects of its
passage. In addition, the studies on the interactions among matter, particles and
radiations allowed, since the X--rays discovery, the
development of today necessary diagnostic tools. The accession of new
detectors, along with the extraordinary development of electronics,
allowed us to obtain fundamental results in medicine: just think about
the positron emission tomography (PET), the digital radiography, the
free elctrons laser \cite{MEDICINE}.

From the moment we decide to use these artificial ``fingers'' in order 
``to touch'' a world out of our ordinary experience, questions of the utmost
importance arise. For example, what is the logical correlation between
the lightning in a scintillator and the passage of a proton?  Is it
possible to gain informations from this lightning? What kind of
informations?  After all, the most important question is: what do we
measure, what do we observe?

The literature on this theme is practically boundless, because of
the importance of these studies in modern physics. We are not trying
to make a complete analysis of the theme here, for this would need at
least a book: we want to give an outline of the most important stages
of the evolution of the stopping power formula and invite the reader to consult the
wide bibliography for a more accurate investigation. 

It is worth noting that, in order to improve the readability, the symbols used in 
this paper will be listed in the Appendix.

\section{The XIX Century}
Even though this story begins at the dawn of the XX century, its roots
are to be found in that extraordinary period of discoveries and
contradictions that characterized the end of the XIX century. In
1873, the scientific community was upset by the publication of the
\emph{Treatise on electricity and magnetism} by J.C. Maxwell, principally
because not many had the necessary knowledge to understand the work of the great
Scottish scientist. The first experimental evidence of Maxwell's
theories came in 1886-1887 only, by H. Herz, who could make experiments
on the electromagnetic waves propagation. Nevertheless, many questions
remained unsolved: in 1885, J.J. Thompson wrote that he still did not
understand what Maxwell called ``quantity of electricity'' \cite{Bellone}.

For this researches, an extremely prolific sector was that of electric
discharges in gases, principally began by Faraday in 1836 \cite{Faraday}.
However, the first valid results were reached only later, thanks to a
quick development of the vacuum technology and to the invention in 1851
of the first rudimental transformer, the Ruhmkorff coil: the device
allowed to obtain electrical discharges at very high voltage. The
pioneers of this sector were Julius Pl\"{u}cker, Willhelm Hittorf, Eugen
Goldstein and Charles Varley \cite{HISTORY}. 
In 1879, it was Goldstein to
propose the name of \emph{cathode rays} to explain the phenomena that take
place in the electric discharge in extremely rarefied gases. The
question on the nature of these rays appeared immediately particularly
muddled and the physicists of the time were divided between those who
were for the undulatory hypothesis (aether waves) and those who were
more inclined for a corpuscular nature. A further contribution came
from William Crookes, who could make a whirl spin by striking it with
cathode rays \cite{HISTORY}.  
Moreover, he demonstrated that the
action of a magnet could deviate these rays.  Therefore, the balance of
science began to be inclined for the corpuscolar hypothesis, even
though Heinrich Herz's shrewdness played an important role in the
undulatory hypothesis.

The crucial stroke came by Jean Perrin \cite{Perrin} and, above all, by Joseph
John Thomson, director of the Cavendish laboratory in Cambridge. He
demonstrated definitively the corpuscolar hypothesis, by succeding in
measuring the ratio between the mass and the charge of a particle, to
which the name of \emph{electron} was given (it had been minted by G.J.
Stoney some years before) \cite{Thom97}. It is curious to note that, years later
J.J. Thomson's son, George Paget, will become famous with the
demonstration of the electron undulatory properties.

In those years other discoveries conquered the world scene: the X--rays  
were discovered in 1895 by Wilhelm C. R\"{o}ntgen and radioactivity
was discovered in 1896 by Henry Becquerel. Physics began the new
century with two kind of particles, atoms and electrons, and with new
radiations (X--rays, radioactivity) and, above all, with many question
marks.

\section{Rutherford and Bohr}
In Septembr 1895, Ernest Rutherford, a young New Zealander with many
hopes and little money, arrived in Cambridge. Thomson's young assistant
recognized two different kinds of radiations emitted by uranium and
called them $\alpha$ and $\beta$, because of the different absorption in matter. In
the following years, Rutherford continued his researches on the nature
of these rays in Montreal and returned to England only in 1907 for a
chair at Manchester University. In his list of possible researches, 
the diffusion of the $\alpha$--rays was included: this lead to the discovery 
of the atomic nucleus.

However, Rutherford's highest contribution was to change the
linguistic approach to the problem by putting in the foreground
issues as collision, scattering and passage of particles through matter 
\cite{RUTH1}. To this purpose the 1905 
article is particularly interesting: it suggested a theory on the origin of 
the scintillation of zinc sulfide exposed to $\alpha$ radiations. 
Becquerel thought that this was due to the cleavage of crystals bombarded by 
$\alpha$ particles, while Rutherford did not write about a direct effect, but 
about a secondary process, the recombination of ions produced during the
radiation passage.  

In the first edition of his book \emph{Conduction of
electricity through gases}, J.J. Thomson had hinted to some issues
regarding the ionization produced by X--rays and by Bequerel's radiations, 
nonetheless basing himself upon the kinetic theory
of perfect gases (diffusion of a fluid in matter) \cite{Thom03}. In the last
chapter only, he considered the problem of what will happen when a
charged particle is stopped. In the second edition of the book,
published in 1906 \cite{Thom06}, Thomson wrote a first theorical approach to the
scattering of two charged particles, elaborated again in a later
article \cite{Thom10}. Nevertheless, he remained tied to the intuitive
conception according to which the collision produced very small
variations on the projectile trajectory; therefore the projectile passed
through matter as if it were in a tunnel bouncing on the walls
(multiple scattering).

This theory was discussed by Hans Geiger with the help of Ernest
Marsden, a freshman: between 1908 and 1910 they published  together
a series of articles, showing the existence of a diffusion of $\alpha$
particles \cite{GEIGER}, \cite{Geiger09}. Even though the 
particles reflected with angles wider than 90$^{\circ}$ were a small percentage 
(1 over 8000):

\begin{quotation}
\ldots it seems surprising that some of the $\alpha$ particles, as
the experiment shows, can be turned within a layer of $6\cdot
10^{-5}$ cm of gold through an angle of $90^{\circ}$, and even
more \cite{Geiger09}.
\end{quotation}

Therefore, Rutherford's surprise in seeing those results was great:
spurred by those results, the New Zealander scientist published, in 1911,
his famous essay, where he described the theory on the atoms
structure: a nucleus surrounded by electrons \cite{Ruth11}. Moreover, he
explained the reason why $\alpha$ particles were reflected with wide angles,
introducing the concept of a strong nuclear field due to the
concentration of positive electric charge in an extremely small area,
nearly punctiform (dimensions $\sim 10^{-14}$ m).

The year after, a young Danish named Niels Bohr arrived at Manchester, after a
disappointing experience with Thomson at Cambridge. Bohr run across the problem 
of the $\alpha$ rays energy
loss while passing through matter. While he was waiting for some radium to
go on with an experimental job, he was attracted by Charles Galton
Darwin, grandson to the great Charles Robert, who was in Manchester at
the time \cite{Pais}. Rutherford made Darwin study the theory of the
$\alpha$--rays penetration in matter, for Darwin had a degree in Applied
Mathematics \cite{Darwin}. Bohr understood that the difficulties the
mathematician met were due to the two possible hypothesis on the
electrons distribution in the atom that Darwin had treated (in the
volume of a sphere centered in the nucleus or on the surface of that
sphere) and decided to treat the atomic electrons as oscillators. In
the meanwhile, Thomson was carrying on researches in this same field,
considering an atom with free electrons and neglecting the results
obtained in Manchester \cite{Thom12}.

It was Niels Bohr who confirmed the validity of  Rutherford group's
approach, with an article ended and published the following year
when he went back to Denmark \cite{Bohr131}. This essay did not only represent
the first sufficiently wide treatment on particles energy loss while
passing through matter, but it was also the prelude to the famous
quantum theory of the atom, published in the same year \cite{Bohr132}. 
Among the most interesting points Bohr underlined there were: the necessity
of treating the collisions between atoms, $\alpha$ and $\beta$ particles in the
same way (apart from the obvious differencies of mass and charge); the
time of collision and therefore a first link with the electromagnetic
theory of dispersion; the evidence, nearly sure, of the fact that
hydrogen has one electron and helium has two.  In 1915, Bohr published
a second essay where he exposed with greater details the theory of
dispersion and considered the relativistic effects \cite{Bohr15}. 

The particles slowing down in the passage through matter remained an interest of Bohr
for his whole life, so that after the discovery of the nuclear fission
the Danish physicist dedicated himself to the study of the passage of fission
fragments in matter \cite{FISSION}, \cite{Bohr48}. 
His last work was written in 1954 with Jens Lindhard
and regards in particular the capture or loss of electrons by fission
fragments in the passage through matter \cite{Bohr54}.

Bohr's approach is based upon classic mechanics: he considers a heavy
particle that passes at such a distance to interact with one electron
of one atom of the material, initially considered at rest, thus
defining the impact parameter $b$. The importance of this parameter
consists in separating near and distant collisions, that have therefore
a different behaviour \cite{Bohr48}. The collisions at distances $d>b$ can be
treated as electromagnetic excitations of harmonic oscillators charged
in an electric field, uniform in space, due to the passage of the
incident particle. On the other hand, if $d<b$, collisions can be treated as
scattering of free electrons by the incident particle. Choosing $b<<\gamma v/\omega$, 
that corresponds to the limit of the impact  parameter beyond which the energy 
tranfer in not efficient, and $b>>|z|e^{2}/mv^{2}\gamma$, that can be seen as a 
dimension of the scattering centre, the energy lost in the collisions is 
given by the formula:

\begin{equation}
-\frac{dE}{dx}=\frac{4\pi N z^{2} e^{4}}{mv^{2}}\left\{\ln
\frac{1.123mv^{3}}{|z|e^{2}\omega}- \ln
(1-\beta^{2})-\frac{\beta^{2}}{2}+R_{i}\right\}
\label{e:bohr}
\end{equation}

\noindent where $R_{i}$ is the function of the impact parameter, though its
contribution to Eq.~(\ref{e:bohr}) is about 0.01\%, giving therefore an
indication of the poor sensibility of Eq.~(\ref{e:bohr}) regarding the parameter.
Moreover, Eq.~(\ref{e:bohr}) shows problems in extreme cases, when $b\rightarrow 0$
or $b\rightarrow \infty$.

\section{Bethe and Bloch}
For many years Bohr's works were practically unrivalled, though there
was no total agreement on the measures taken, in particular on the $\alpha$
and $\beta$ particles. In the twenties, after the birth of quantum mechanics,
the deficiencies of Eq.~(\ref{e:bohr}) were regarded as the failure of classic
mechanics in the treatment of atomic phenomena and, in particular, to
the fact that energy tranfers, in this kind of reactions, take place for
discrete quantities.

The first news arrived in the thirties, when Hans Bethe faced the
problem again by means of the new quantum mechanics \cite{Bethe30}. 
To tell the truth, there had been other attempts in the twenties, though not
completely satisfactory. Bethe solved the problem by using Born
approximation. This approximation requires that the amplitude of the wave 
scattered by the atomic electron field must be small when compared to the one of 
the incident wave (undisturbed). This is expressed with the condition \cite{Bethe53}: 

\begin{equation}
\frac{ze^{2}}{\hbar v}<<1
\label{e:born}
\end{equation}

\noindent Condition Eq.~(\ref{e:born}) is satisfied in case of high speed and low electric 
charge of the incident particle. A further simplification came when
the speed of the incident particle is considered to be much higher than that of the
electrons of the absorbent material atoms:

\begin{equation}
	E>>\frac{M}{m}E_{el}
	\label{e:velo}
\end{equation}

\noindent where $E$ is the energy of the incident particle and $E_{el}$ is the
ionization potential of the atomic electron.
Under these conditions and considering non-relativistic speeds the
energy loss is:

\begin{equation}
	-\frac{dE}{dx}=\frac{4\pi N z^{2} e^{4}}{mv^{2}}B
	\label{e:bethe1}
\end{equation}

\noindent that gives the stopping power of the material. The term $B$
(\emph{stopping number}) is expressed as:

\begin{equation}
	B=Z\ln \frac{2mv^{2}}{I}
	\label{e:bethe2}
\end{equation}

Comparing Eq.~(\ref{e:bohr}) to Eq.~(\ref{e:bethe1}) and Eq.~(\ref{e:bethe2}) we note that the 
logarithmic term in Bethe's formula does not depend from the incident particle charge.
In the case of relativistic speeds, Eq.~(\ref{e:bethe2}) becomes 
\cite{Bethe32}:

\begin{equation}
	B=Z \left\{\ln \frac{2mv^{2}}{I}-\ln 
	(1-\beta^{2})-\beta^{2}\right\}
	\label{e:bethe3}
\end{equation}

It is worth noting that Bethe's and Bohr's formulas have a field of
validity each: therefore they can be considered complementary. A
synthesis of the two formulas was carried out in 1933 by Felix Bloch \cite{BLOCH},
who obtained:

\begin{equation}
	-\frac{dE}{dx}=\frac{4\pi z^{2} e^{4}}{mv^{2}}NZ \left\{\ln 
	\frac{2mv^{2}}{I}+\psi(1)-\mathrm{Re}\psi\left(1+i\frac{ze^{2}}{\hbar 
	v}\right)\right\}
	\label{e:bloch}
\end{equation}

\noindent where $\psi$ is the logarithmic derivative of Euler's $\Gamma$ function. 
Eq.~(\ref{e:bloch}) becomes Bethe's formula for 
$ze^{2}/\hbar v=0$, while it is reduced to Bohr's formula when $ze^{2}/\hbar v>>1$. 
It is worth noting that Eq.~(\ref{e:bloch}) is referred to
the case of non--relativistic speeds. For $\beta\rightarrow 1$ it is sufficient to 
detract the terms $[\ln (1-\beta^{2})+\beta^{2}]$ to the quantity enclosed 
within brace brackets

In 1953, a detailed work by Bethe and Ashkin \cite{Bethe53} was added to the
excellent review done by Bohr \cite{Bohr48}; the new work included
also various cards and graphs relatives to experimental measures
and it analysed widely any  possible anomaly.

\section{Today}
What is today known as the Bethe-Bloch formula is the principal
instrument to reckon the energy loss due to particles passage through
matter. Several correction coefficients were added to
consider many effects \cite{REVIEW}, 
nevertheless the Bethe-Bloch formula still remains the starting point.  

A first correction was signaled by Bhabha \cite{Bhabha} 
and regarded the maximum energy that can be transferred to an electron,
initially at rest, in a free collision, that results:

\begin{equation}
	T_{max}=\frac{2mc^{2}\beta^{2}\gamma^{2}}{1+2\gamma\frac{m}{M} 
	+(\frac{m}{M})^{2}}
	\label{e:bhabha}
\end{equation}

\noindent and therefore takes the place of the simple $2mv^{2}$ in equation 
Eq.~(\ref{e:bethe2}).

Another important corrective factor is the so called ``density effect'',
due to the fact that the electric field of the incident particle has
a tendency to polarize atoms. As a matter of facts, Bohr, Bethe and
Bloch's analysis regarded gases at low density; these gases can be seen
as an aggregate of independent particles, which does not happen in
solid bodies or in liquid ones.
In the fourties, Enrico Fermi noticed that the energy loss did not
depend from physical parameters only, those parameters Bethe and
Bloch took account of, though it depends also from the material
dielectric properties \cite{Fermi}. 
It is well worth to remember that in
1924 Fermi had already noticed the analogy between the collision and
dispersion phenomena and had proposed to evaluate the stopping power 
by basing himself upon the empirical evidence of the
high frequency radiation absorption \cite{Fermi24}.
To evaluate the density effect Fermi used classical electrodynamics and a
simplified model of the atom; later on, Gian Carlo Wick \cite{Wick} 
and, independently (the time gap was due to the war), 
Otto Halpern and Harvey Hall \cite{Halpern}, 
used a more complicated model, a multifrequencies one that takes account of the
dispersion;  moreover Halpern and Hall considered also the damping of
the conduction electrons. An empirical
expression of the density correction for a wide range of materials
was given by Sternheimer \cite{Stern1}. Later on, Sternheimer himself along with
Peierls, gave a generalized version of it \cite{Stern2}.

However, the classical and semiclassical approach to density effect was readily
substituted by the quantum one \cite{Fano}, \cite{Fano63}, even though it did 
not disappear completely. In 1960, Landau and
Lifshitz still gave a semiclassic version of it in their famous \emph{Course
of Thoretical Physics} \cite{Landau} and in 1980 Ahlen wrote that this kind of
approach may still be satisfactory, if it is used in energy intervals
where polarization effects are not prevalent \cite{REVIEW}. Between the
macroscopic effect and the microscopic one, another way has to be
signaled; it was followed by Aage Bohr to try and evaluate the
relativistic effects, above all \cite{ABohr}.
These are the cause of discordance between the classic density
correction and the quantum one: as a matter of fact it is necessary
to make a distinction between a longitudinal contribution and a
transversal one for the first only is present in the classic derivation
\cite{Fano}. The transversal term puts on a significant 
probability at relativistic speeds and is to be held responsible for \v{C}erenkov
radiation.

Another problem in the Bethe-Bloch formula is made up by the mean excitation
potential of the absorbing material ($I$), which is
function of $Z$. Being it an average value it is possible to mantain the
computation at a moderate accuracy level, which is an advantage. On the
other hand, in materials composed by various elements and in particular
in organic materials, the chemical bond and other aggregation
characteristics may insert sensitive variations through changes in the
spectrum of the excited levels \cite{Fano63}.

Another assumption effectuated in the Bethe-Bloch formula is that the
incident particle is decidedly faster than atomic electrons as
specified in equation Eq.~(\ref{e:velo}). Nevertheless, as the incident particle speed
decreases, the contribution to the stopping from the inner shell electrons
($K,L,\ldots$) disappears. It is therefore necessary to insert a further
correction coefficient to take this phenomenon into account \cite{Bichsel}, 
\cite{Fano63}.

Moreover, there may be differences in conformity with the
incident particle, if it is positive or negative. It is the so called
``Barkas effect'', discovered in the 50's studying the decay of $K\rightarrow 3\pi$: 
it was observed that the $\pi^{-}$ particles had slightly longer paths than the 
$\pi^{+}$ ones \cite{Barkas}. A further correction coefficient is then
introduced and it depends, according to the particle, on $z^{3}$ or $z^{4}$ 
\cite{Lindhard76}.

For other correction coefficients or changes to be introduced in the
Bethe-Bloch formula according with its use, we would suggest to read
the wide reviews listed in the bibliography, in particular \cite{Bethe53}, 
\cite{REVIEW}.

Lately, the only news of a certain prominence is  the theory by Jens
Lindhard, a Bohr's scholar, and Allan S\o rensen, relative to the energy
loss of heavy ions at relativistic speeds; this theory starts from
the well--known Bethe-Bloch formula \cite{Lindhard96}. The authors notice that 
the deviations that take place in the case of relativistic ions are to be expressed
in terms of cross section for the scattering of an electron, free from
the Coulomb field of ions; the authors carry out the necessary
computations using Dirac equation for relativistic ions.

\section{Conclusions}
In 1998, the \emph{Particle Data Group} relates this expression \cite{REVIEW}:

\begin{equation}
	-\frac{dE}{dx}=\frac{4\pi Nr_{e}^{2}mc^{2}z^{2}Z}
        {A\beta^{2}}\left[\frac{1}{2}\ln\left(\frac{2mc^{2}
        \beta^{2}\gamma^{2}T_{max}}{I^{2}}\right)-\beta^{2}
        -\frac{\delta}{2}\right]
	\label{}
\end{equation}

\noindent which is adaptable to the majority of cases. Some correction
coefficients - already discussed above - follow.
As we can see the Bethe-Bloch formula is sill today a ``finger'' of
physics, perhaps the most important one. Maybe, it is a botched up
one but is is still very efficient and it is a necessary instrument not
only for the understanding of the world around us though also for its
technological applications. It is well worth to think about the use of
radiations in cancers treatment, a therapy that requires a great
caution to avoid  telling the famous joke: ``the operation was
successfull, but  the patient died''.

\section{Acknowledgements}
A sincere thank to Federico Palmonari for the critical revision of the text
and for his useful suggestions. A special thank to the Library of the Department of
Physics (University of Bologna), without which this research would not be possible.

\section*{Appendix}
List of symbols:\\
$Ze$: electric charge of the absorbing material;\\
$ze$: electric charge of the incident particle;\\
$A$: atomic mass of the absorbing material;\\
$M$: rest mass of the incident particle;\\
$m$: electron rest mass;\\
$r_{e}$: electron classical radius;\\
$v$: incident particle speed;\\
$c$: speed of light in vacuum;\\
$\beta=v/c$;\\
$\gamma=(1-\beta^{2})^{-1/2}$;\\
$\delta$: density correction;\\
$N$: number of electron for absorbing material volume unit;\\
$d$: distance;\\
$b$: impact parameter;\\
$I=\hbar\omega$: mean excitation energy of absorbing material.\\

\end{document}